\documentclass[conference]{IEEEtran}
\pdfoutput=1 
\usepackage{url}
\usepackage{eurosym}
\usepackage{todonotes}
\usepackage{footnote}
\usepackage{subfigure}
\usepackage{balance}
\usepackage{colortbl}
\usepackage{amsmath}
\usepackage{bm}
\usepackage{mathpazo}
\usepackage{paralist}    
\usepackage{colortbl}
\usepackage[normalem]{ulem}

\definecolor{equ}{rgb}{0.93,0.93,0.93}

\begin{document}



\title{Wi-Fi Offload: Tragedy of the Commons or Land of Milk and Honey?}






















\author{\IEEEauthorblockN{Patrick Zwickl\IEEEauthorrefmark{1}, Paul Fuxjaeger\IEEEauthorrefmark{1}, Ivan Gojmerac\IEEEauthorrefmark{2}, Peter Reichl\IEEEauthorrefmark{2}\IEEEauthorrefmark{3}}
\IEEEauthorblockA{\IEEEauthorrefmark{1}FTW Telecommunications Research Center Vienna, Donau-City-Stra\ss e 1/3\textsuperscript{rd} floor, \\ 1220 Vienna, Austria, E-mail: \{zwickl$\vert$fuxjaeger\}@ftw.at}
\IEEEauthorblockA{\IEEEauthorrefmark{2}University of Vienna, Faculty of Computer Science,  W\"ahringer Stra\ss e 29, \\
1090 Vienna, Austria, E-mail: \{ivan.gojmerac$\vert$peter.reichl\}@univie.ac.at}
\IEEEauthorblockA{\IEEEauthorrefmark{3}Universit\'{e} Europ\'{e}enne de Bretagne/T\'{e}l\'{e}com Bretagne, 2, rue de la Ch\^{a}taigneraie, \\ 35576 Cesson S\'{e}vign\'{e}, France}
}

\maketitle

\begin{abstract}
Fueled by its recent success in provisioning on-site wireless Internet access, Wi-Fi is currently perceived as the best positioned technology for pervasive mobile macro network offloading. However, the broad transitions of multiple collocated operators towards this new paradigm may result in fierce competition for the common unlicensed spectrum at hand. In this light, our paper game-theoretically dissects market convergence scenarios by assessing the competition between providers in terms of network performance, capacity constraints, cost reductions, and revenue prospects. We will closely compare the prospects and strategic positioning of fixed line operators offering Wi-Fi services with respect to competing mobile network operators utilizing unlicensed spectrum. Our results highlight important dependencies upon inter-operator collaboration models, and more importantly, upon the ratio between backhaul and Wi-Fi access bit-rates. Furthermore, our investigation of medium- to long-term convergence scenarios indicates that a rethinking of control measures targeting the large-scale monetization of unlicensed spectrum may be required, as otherwise the used free bands may become subject to tragedy-of-commons type of problems.










\end{abstract}





%
\IEEEpeerreviewmaketitle


\section{Introduction}

The number of broadband connections in urban industrialized areas has grown massively in recent years\footnote{OECD: \url{http://www.oecd.org/sti/broadband/oecdbroadbandportal.htm}, last accessed: June 16, 2013}. The achievable link-rates of Cable/DSL/FTTH-based fixed lines have steadily increased and at the same time most Customer-Premises Equipments (CPEs) have been gradually upgraded to include 2.4/5 GHz IEEE~802.11n frontends, making them capable to downstream-rates in the order of 100Mbit/s in the high SNIR region. Recent empirical studies, e.g., \cite{FuxjagerFGR10}, have shown that in densely populated metropolitan regions the wireless coverage areas of Wi-Fi enabled CPEs do already spatially overlap. The measurement study presented in \cite{FuxjagerGFR11} also illustrates that a dominant fixed-line broadband operator would easily be able to aggregate its CPEs to form a complete radio access network covering large portions of the urban environment. The disruptiveness of this approach for mobile radio access network operators, however, becomes evident when considering backhaul limitations: even if all cellular sites in a given area were upgraded to radio access technologies with very high spectral efficiency (LTE eNodeBs with multiple sectors and maximum channel bandwidth), the aggregated per area throughput will ultimately be limited by the respective \emph{backhaul capacity}. While its performance limit can be further pushed by gradually deploying dedicated high-capacity lines, the operational cost of backhauling by aggregation of a multitude of broadband connections in the same area will remain strictly strictly lower.


Through the recent development of automatic client-side network-access selection capabilities the attractiveness of automatic high-volume Wi-Fi offload is raised. The majority of current smartphone/tablet platforms already support recent standard amendments such as IEEE~802.11u and Hotspot~2.0 as well as functionalities to use SIM-based authentication (EAP-SIM) mechanisms. These advancements are likely to lead to an opportunistic use of Wi-Fi networks whenever possible, which may effectively eliminate remaining barriers at the users' side.





The aforementioned densification together with the reduction of access-barriers has significantly narrowed the gap between cellular and IEEE~802.11-based networks, which may potentially be disruptive for the current radio access network industry. But which role will densified LTE networks play in this respect? In other words, what are the weaknesses of Wi-Fi access in comparison to LTE small-cell networks? Seamless inter-system handover which offers session continuity is one feature that Wi-Fi is not able to deliver with the same ease as cellular access -- but -- the expected growth in data volume is mainly driven by services that correspond to nomadicity, for which session continuity is not required\footnote{Cisco Visual Networking Index (2013): \url{http://www.cisco.com/en/US/solutions/collateral/ns341/ns525/ns537/ns705/ns827/white_paper_c11-520862.pdf}, last accessed: July 3, 2013}. Furthermore, application service providers have started to deploy mechanisms to cope with IP-address changes and short outages in non-interactive services. This means that the relevance of seamless handover has been reduced, while current cellular networks still center their underlying architecture on the realization of mobile voice services. Such a focus also leads to another differentiation between the two domains, i.e., in terms of complexity and the cost of client-side equipment. On the Bill-of-Materials (BoM), an LTE baseband chip is up to five times more expensive\footnote{\url{www.isuppli.com/teardowns/}, last accessed: June 16, 2013} than an IEEE~802.11 transceiver of the latest generation. This price difference and the customers' fear of additional running expenses due to SIM contracts effectively disadvantage cellular access over Wi-Fi in the long run.


In this paper, we argue that licensed small-cell networks \emph{will not become relevant} in cases where the system is noise-limited, which is the case when the physical environment compartmentalizes the medium into units without significant interference coupling. The major advantage of licensed access is that the hierarchical protocol stack and the exclusive-use option allows to operate an efficient radio resource coordination mechanism, which is in fact also strictly needed when the spatial orthogonality is \emph{not given anymore}.
However, spatial overlap in the unlicensed spectrum is already observable in certain densely populated regions. This is especially noteworthy since many fixed \emph{and} fixed/mobile operators who are active in European and North American metropolitan regions have started to push and roll-out Wi-Fi offload strategies in recent months: SFR and Free in France, Comcast, Deutsche Telekom, Belgacom, just to name a few. This paper is, therefore, dedicated to the analysis of tensions arising from technological and business convergence of unlicensed and licensed spectrum.


The remainder of this work is structured as follows: Section~\ref{sec:rw} provides an overview of related work regarding mobile network offload business and technology, after which in Section~\ref{sec:model} we analytically study the formation of unexpected market convergence equilibria and dominant strategies. By scaling the problem scope to a market dimension, we discuss a series of resulting Wi-Fi market challenges in Section~\ref{sec:market}. Finally, in Section~\ref{sec:outcome} we point out a series of realistic outcome scenarios and conclude the paper with some final remarks in Section~\ref{sec:concl}.
 
\section{Related Work} \label{sec:rw}


Due to increasingly overloaded mobile 3G networks, the offload of non-realtime traffic has recently raised significant academic interest \cite{han10}. Although manual traffic offload has de facto been a success story (for up to 65\% of the relevant university traffic \cite{lee10}), Wi-Fi offload still faces clear obstacles in some scenarios due to Wi-Fi reception challenges of current generation smartphones---according to \cite{liu12}, the majority of student traffic is still in the 3G network. Thereby, \cite{lee10} argues that proper utilization of Wi-Fi networks may even considerably contribute to energy savings, which underlines the ecological desirability of consistent Wi-Fi offload.


For the automation of traffic offload, a feasible selection of traffic seems to be required. Siris and Kalyvas \cite{siris13} have followed up on this idea by offloading delay tolerant traffic based on a designed mobility prediction. Thus, a concentration on nomadic users for mobile traffic offload, e.g., in home scenarios, combined with mass deployment of accessible Wi-Fi hotspots may be of highest interest. In addition, application-awareness seems to be necessary in order to optimally target user requirements, e.g., through Quality of Experience (QoE) trials for applications which have important delay constraints\footnote{``Apple Not Throttling iPhone or iPad Cellular Throughput via Carrier Bundles'': \url{http://www.anandtech.com/show/7037/apple-not-throttling-iphones-ipads-cellular-throughput-via-carrier-bundles-}, last accessed: June 16, 2013}. 



Differentiated by the following list of driving actors, Casey et al. \cite{casey10} have compiled a set of Wi-Fi/Femto offload Value Network Configurations (VNCs): mobile operators, broadband access operators (fixed line), combined fixed and mobile operators, the access aggregator role (e.g., FON\footnote{\url{http://www.fon.com/}, last accessed: June 14, 2013}), service providers and device manufacturers, as well as site/venue owners. Investigating market convergence effects, the present paper will concentrate on the fixed/mobile operator and access aggregator cases. These VNCs have been elaborated by \cite{casey12} in order to illustrate two-sided market issues towards a successful market evolution, e.g., w.r.t. required platform subsidies. Cost aspects for offloading base station traffic to femto cells have specifically been targeted by \cite{gronsund13}. However, to the best of our knowledge, the conflict of business interests and market challenges involved in offloading traffic from mobile networks to unlicensed spectrum Wi-Fi solutions has not yet been targeted in literature.

\section{Model} \label{sec:model}

We will now start to analytically assess the strategic convergence of network offloading from cellular networks to Wi-Fi networks (utilizing unlicensed spectrum and fixed line Internet access) by game-theoretically investigating locally concentrated competition under weak control mechanisms (as currently practiced in the unlicensed spectrum environment). This section is intended to illustrate the need for finding cooperative means for the large-scale monetization of the unlicensed spectrum.



\begin{figure}[htbp]
\centering
\setlength\fboxsep{2 pt}
\setlength\fboxrule{0.5pt}
\fbox{
\includegraphics[width=0.45\textwidth]{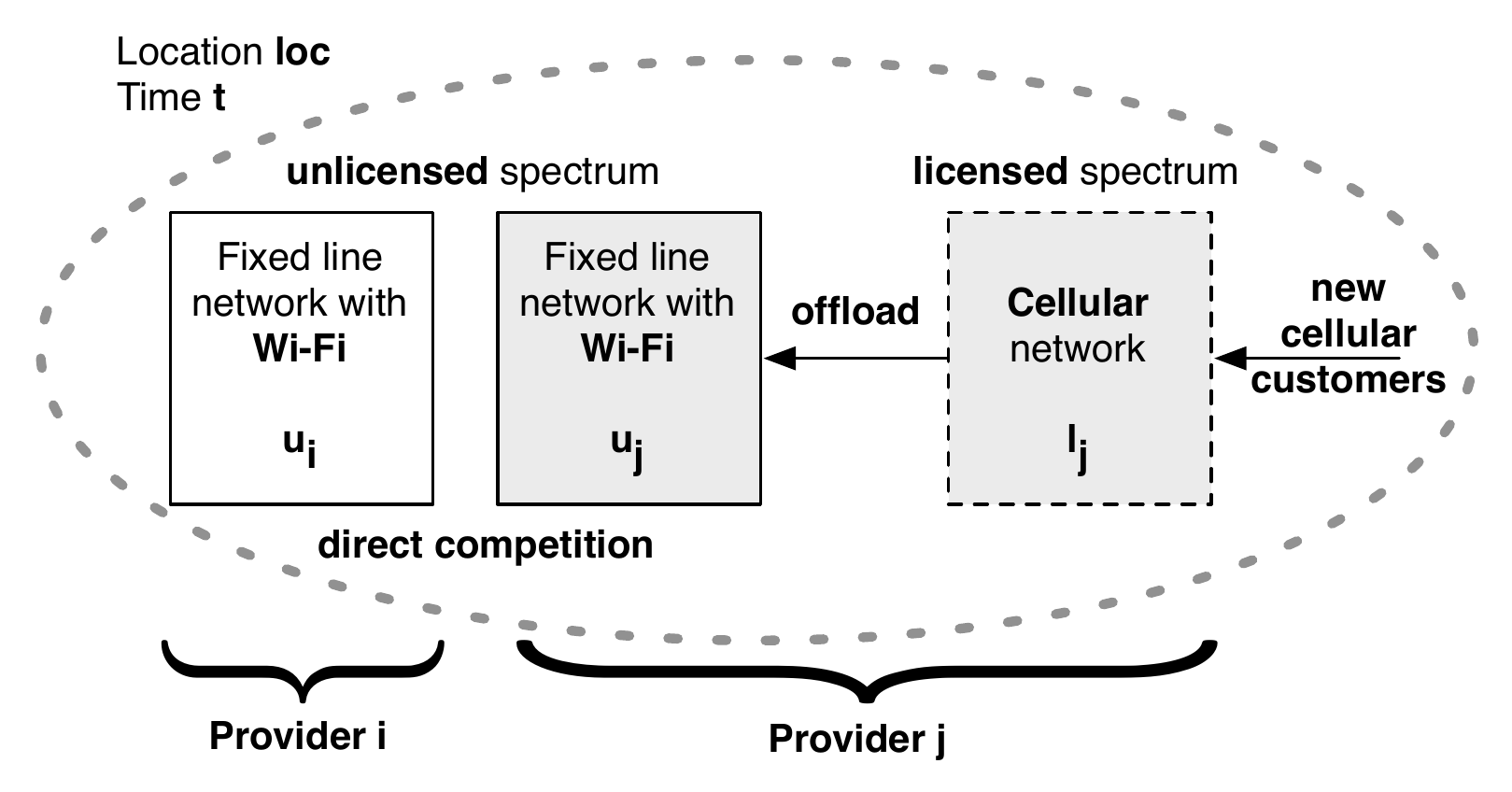}
}
\caption{A competitive market with two competing providers where one is offloading.}
\label{fig:bigpicture_game}
\end{figure}

Let us assume there are two competing Internet Service Providers (ISPs) in location $loc$ at point in time $t$ as depicted in Figure~\ref{fig:bigpicture_game}: Provider $i$ provides only wireless Internet service in the unlicensed spectrum network $u_i$ (backed by the fixed line backbone), while its competitor $j$ in addition also provides cellular Internet services in the licensed spectrum, i.e., $u_j$ and $l_j$. For simplification reasons we assume both are present at the same market/location $loc$ at time $t$.

We further model the QoS $q(d, C)$ sensitive to traffic demands $d$ and the available capacity $C$\footnote{In the case of Wi-Fi, $C$ represents the physical  capacity minus intra-customer site traffic.} (i.e., the bigger $d$ for a given network, the higher the congestion)---comparable to \cite{sigcomm13}. Only for the unlicensed spectrum case the overall capacity is shared, thus in analogy to \cite{shetty2008modeling} the $q \in [0,1]$ can be described as

\begin{equation} 
\label{eq:qos}
q(d^u_i + d^u_j, C^u) = 1 - \frac{d^u_i + d^u_j}{C^u}
\end{equation}


where the joint capacity $C^u$\footnote{Wi-Fi spectrum being unassigned to specific providers or users} is static in $t$ and the demand $d$ satisfies $d^u_i + d^u_j \leq C^u$---conversely $d^l_j \leq C^l_j$ for the licensed spectrum. We model the demand $d$ as function $d^{\{u,l\}}( \Delta p, \Delta q)$ which is decreasing in ``price distance'' $\Delta p$ and increasing in $\Delta q$, where the terms $\Delta p$ and $\Delta q$ are defined as relative price and QoS tradeoff over the next best alternative\footnote{Current contract serves as reference point, while absolute access fees may also limit the user exclusion rate.}, i.e., loyalty. Assuming the user is considering an active changeover from $l_j$ to $u_j$, i.e., replacing a cellular with a fixed line contract, these terms can be described by


 
 

$$\Delta p = p^l_j - p^u_j \quad , \quad \Delta q = q^l_j - q^u_j \quad .$$

In turn, we can specifically derive the revenue of provider $j$ for satisfying $d$ from 


 
 \begin{equation} 
\label{eq:revenue}
\pi^n_k(d^n_k) = d^n_k p^n_k = d^n_k(\Delta p, \Delta q) p^n_k \quad, 
\end{equation}

where $p^n_k$ describes the price charged by provider $k$ which is of type $n$, by the sum of revenues in $u_j$ and $l_j$,





 \begin{equation} 
\label{eq:sumrevenue}
\pi^u_j(d^u_j) + \pi^l_j(d^l_j) \quad,
\end{equation}

and conversely for $i$'s unlicensed spectrum network.

Thus whenever provider $j$ starts to offload traffic, the shared spectrum is challenged---according to \eqref{eq:qos} the cellular network attractivity in terms of $QoS$ and $p$ relatively rises on the expense of Wi-Fi networks. From the formulation of $d$ we can infer that the demand will shift towards relatively better alternatives (as well as the opposite), if available---i.e., the prospective demand of $j$'s cellular network increases with its improved QoS or decreasing QoS of $i$'s network.

Let us also assume that the cost per bit per Hertz\footnote{This encompasses spectrum fees and backhauling costs. Also note that real-world investment cost curves for meeting capacity demands are discrete rather than continuous. However, we assume the control over the degree of offloading (especially w.r.t. locally available resources) is coarse-granular, thus costs per bit $E$ are a feasible metric.} $E$ satisfies $E^l_j > E^u_j$, which may be explained by the immanent resource scarcity in cellular networks (limited excess resources due to limited available spectrum) in addition to high license fees. In practice, the price $p^l$ for cellular network customers is higher than for its unlicensed spectrum counterpart $p^u$ (Wi-Fi)---thus, $\pi^{l}(d') > \pi^{u}(d')$ for any given $d'$.


Due to offloaded cellular traffic, capacity in $l_j$ is freed up to be resold to new cellular customers, which may impair the QoS of $i$ and $j$'s Wi-Fi network. Under these conditions a dominant incentive to offload traffic from the cellular network $l_j$ to $u_j$ exists for player $j$ as long as $QoS(d^u_j) > QoS(d^l_j)$ holds and $p$ and thus $\Delta p$ remains static\footnote{Unaccounted traffic may slightly lower the price for the end customer depending on the specific charging regime}. This is explained by the revenue prospects of reselling freed capacity, the lower cost per bit $E$ in the fixed line, and unimpaired demand $d^l_j$ and thus $\pi^l_j$ from existing customers which are not aware of these offloading practices, while the revenue- and demand-effective \emph{externality } of decreasing QoS in the unlicensed spectrum is equally shared between $i$ and $j$. Beyond that point, offload is still dominant whenever cost reductions, $E^l_j-E^u_j > 0$, together with revenues from reselling freed capacity still exceeds the demand and revenue loss in $u_j$.




By utilizing application-aware offloading strategies, e.g., isolation of sensitive traffic types, a second dominant incentive exists to offload as much low priority bulk traffic as possible, while sensitive traffic, e.g., video conferences, is kept in the cellular network. Thus, selective offloading may even be feasible if $QoS(d^u_j) \leq QoS(d^l_j)$.




Due to the higher costs and the unique selling point of high quality cellular networks, quality optimization efforts of ISPs will be centered on these networks. Thus, when $d^u_i + d^u_j + d^l_j$ represents the constant total demand at $\langle loc, t \rangle$ and $QoS(d^u_i) \cong QoS(d^u_j)$ due to shared spectra, then there exists an incentive for $j$ to assure $QoS(d^l_j) > QoS(d^u_{i/j})$. While in the unshared spectrum the loss of quality and demand is shared, the full revenue potential can be targeted by $j$'s licensed spectrum network $l_j$.

Summarizing what has been said for intra-provider offloading strategies, offloading is dominant whenever one of the following conditions is satisfied: (i) $\Delta q_{j} < 0$, (ii) $\pi^l_j$ or $C$ yields outperform $\pi^u_j$ reductions, (iii) bulk traffic is offloaded without the customer being necessarily aware of it.



Therefore, we apply this model to a system with two application types $z=\textit{\{b, v\}}$ ordered by increasing quality demands, i.e., lowest quality $q$ requirements for bulk traffic $b$ as well as  premium traffic in the form of highly delay-sensitive video conferences $v$. Then an \emph{intra-provider equilibrium} demand ${d}^u_j*$ results whenever for any marginal piece of bulk traffic provider $j$ is indifferent between offloading to $u_j$ or retaining in $l_j$:



\begin{equation} 
\label{eq:intradomaineq}
l_j \rightarrow u_j: \Big( q^u_j(d^u_j`) < \underline{q_{z=b}} \Big) \wedge \Big( q^u_j(d^u_j) > \underline{q_{z=b}} \Big)
\end{equation}
\begin{equation} 
\label{eq:intradomaineq}
\implies q({d}^u_j*) = \underline{q_{z=b}} \quad, 
\end{equation}


where $d^u_j` = d^u_j + \epsilon$ for some $\epsilon > 0$. Thus, a convergence around the minimum quality requirements $\underline{q_{z=b}}$ for bulk traffic, i.e., high utilization of $u_{j}$, is inferred, which reflects the conditions (i) to (iii).









Beyond that, the sharing of the freely available Wi-Fi spectrum, i.e., a common good, also postulates a competition over resources between $i$ and $j$. Corresponding to $j$'s intra-domain equilibrium, $i$'s best response is high rather than low utilization, i.e., lowering utilization shifts revenues from $i$ to $j$, as $q^u$ rises and can be traded for $q^l_j$. Due to higher $p$ and thus revenues, both $i$ and $j$ strive for increasing the volume of carried $v$ traffic.

A second stage \emph{inter-provider equilibrium} thus exists, whenever for a marginal piece of premium traffic $z=v$ provider $j$ becomes indifferent between keeping traffic in $u_j$ or $l_j$ (in analogy to \cite{Gibbens:2001ty}):

\begin{equation} 
\label{eq:intradomaineq}
\Big( q^l_j(d^l_j``) < \underline{q_{z=v}} \Big) \wedge \Big( q^l_j(d^l_j) > \underline{q_{z=v}} \Big)
\end{equation}
\begin{equation} 
\label{eq:intradomaineq}
\implies q({d}^l_j*) = \underline{q_{z=v}}  \quad, 
\end{equation}

where $d^l_j`` = d^l_j``(z = b) + d^l_j``(z = v)$ and
\begin{align}
d^l_j``(z = v) &= d^l_j(z=v) + \epsilon \quad , \nonumber \\
d^l_j``(z=b) &= d^l_j(z=b) - \epsilon` \nonumber
\end{align}

 for some $\epsilon > 0$ and $\epsilon` \geq 0$, as well as sufficient premium demand $d^l_j(z=b)$ exists and $\pi^l_j(\delta) > \pi^u_{i,j}(\delta)$, as the overall demand is constant. Thus, $j$ has no incentive to provision a QoS level above $\underline{q_{z=b}}$, thus $q(d^u_{i,j})$ will settle at $\underline{q_{z=b}}$ due to \eqref{eq:intradomaineq}. Thus, $i$'s dominate strategy is to materialize all possible bulk traffic, as premium traffic under resource scarcity in $\langle loc, t \rangle$ will be captured by $l_j$.

While in the first phase, the efficient freeing up of cellular network capacity through Wi-Fi offloading (also raising the quality level) may be attractive (and dominant), the control over the unlicensed spectrum may become of strategical relevance. On the one hand, the provider best utilizing the available unlicensed spectrum will profit from capacity gains (increased revenue due to higher demand and more capacity\footnote{Probably also higher prices due to increased demand.})---an equilibrium at high utilization of all providers and yielding a moderate QoS will exist (as long revenue effective or reducing costs). As a consequence, this leads to the \emph{tragedy of the commons} where \emph{externalities } caused by inconsiderate use of common resources are ignored. On the other hand, today's fixed line operators are actively striving towards Wi-Fi services (due to customer interest) and may aim at complementing cellular network services. In this light, provider $j$ may improve its strategical positioning by overloading shared spectrum in order to raise the relative QoS advantage of their cellular network services. In saturated markets, strategic optima may hence converge towards the blockage of shared spectrum rendering the position of combined cellular and fixed line operators strategically beneficial. Note that in practice, a full utilization of shared spectrum in $\langle loc, t \rangle$ and especially in a whole market may not be fully synergistically controllable and competition in both unlicensed and licensed spectrum may avoid such a convergence\footnote{This may be compared to business strategies of artificial scarcities under collusion or cartels.}.


With the sketched mechanism a simple two class system is established ($l_j$ as premium and $u_j$ as standard class), where additionally premium services keeping all bulk traffic of a customer in $l_j$ may be sold.


\section{Market Scale} \label{sec:market}



Loosely pairing the implications of our model with market-level considerations, a natural competition between one provider's own unlicensed and licensed spectrum wireless Internet offers may arise. Considering the upcoming \emph{unconstrained bandwidth} offers in cellular Internet access, as currently appearing in Europe\footnote{``3~SuperFlat'':~ \url{http://www.drei.at/webshop/prepareSelectionTarifDetail.do?tarifId=ATS0076&snId=C72633}, last accessed: June 14, 2013}, combined with an efficient offloading mechanism (e.g., directly accessing Wi-Fi access points of other unlicensed spectrum customers), fixed line access contract may appear redundant to many customers, as long as the alternative service coverage is acceptable. This, may lead to oscillations between states of abandoning fixed line access and re-deploying fixed access points due to reduced coverage or service quality.


As an interesting case in Wi-Fi offload scenarios, customers situated in densely populated areas or close to points of interest may provide Wi-Fi access to a broad range of visitors who would previously have been using licensed spectrum roaming, whenever upcoming global agreements, e.g., with FON, turn into practice. This may on the one hand impair the customer's QoS through limited wireless or even backhaul resources, unless dedicated primary-customer prioritization techniques are applied. On the other hand, paying for Internet access may also appear to be unjustified or unfair to the primary customers sharing their connectivity. Therefore,  compensation schemes may become necessary in order to reach an equilibrium in which both reasonable coverage as well as perceived fairness are achieved for the vast majority of customers involved. 


Active sabotage through unlicensed spectrum overload may be an interesting extremum in weakly competitive markets facing greedy incumbent players or monopolists. Thus, while no customer of their own is present\footnote{This would necessitate smart access point devices at customer sites}, sabotaging may artificially limit the capabilities of Wi-Fi.


\section{Outcome Scenarios} \label{sec:outcome}



The present section will outline three realistic medium- to longer-term convergence scenarios being inferred from the discussed market challenges in Section~\ref{sec:market}. Above all, we will differentiate by cases where for a given tuple $\langle loc, t \rangle$ sufficient or insufficient high-speed Internet access backhaul resources $C^b$ are available to meet the Wi-Fi capacity $C^u$ at $\langle loc, t \rangle$.




For the case where $C^b \geq C^u$, the capacity sharing may result in two distinct equilbria:


\begin{enumerate}
    \item \textbf{Self-balancing}: The usage of the Wi-Fi network will reach a stable equilibrium after a series of transitions between the fixed and cellular Internet access technologies. In the optimal case, the Wi-Fi quality is fostered through an improved or even optimal distribution of access points. 
    \item \textbf{System deadlock}: The Wi-Fi network as strategic resource is intentionally or unintentionally impaired or destroyed. This may be caused by cellular operators attempting to artificially increase the value of their licensed spectrum resources through excessive Wi-Fi offload (\emph{tragedy of the commons}), or by churning fixed Internet customers who see themselves treated unfairly with respect to the value they add to and receive from their commercial offerings (due to Wi-Fi offload), thus reducing the global availability/density of Wi-Fi access points.
\end{enumerate}


For the opposite case, i.e., $C^u > C^b$, another convergence point is of interest:


\begin{enumerate}\setcounter{enumi}{2}
    \item \textbf{Backhaul-based limitation}:
 Whenever the access bandwidth aggregate across multiple operators generally exceeds the available Wi-Fi capacities in common scenarios (e.g., an urban setting with typical access link / Wi-Fi placement density), there will be no formation of a tragedy of the commons-situation. In other words, Wi-Fi capacity does not represent the limiting factor in terms of bandwidth, rendering the wireless last hop transparent in end-to-end available bandwidth considerations for the individual users.
\end{enumerate}

The distinction between $C^u$ being greater than $C^b$ or vice versa is dependent on factors that all strongly depend on the location and on the given backhaul infrastructure, which needs to be systematically understood in further work.

\section{Conclusions} \label{sec:concl}



Whenever capacity is scarce, combined unlicensed spectrum (Wi-Fi) and licensed spectrum (cellular network) providers tend to economically advance (revenue, market powers, etc.) on the expense of Wi-Fi-only providers. The present work has been able to illustrate equilibria shifting premium traffic to cellular networks and bulk traffic to highly utilized Wi-Fi networks. Thus, while cellular networks may constitute quality as unique selling point, the revenues of Wi-Fi networks transitively face a localized endangerment. By scaling these considerations to a market dimension, we have been able to find indicators for phenomenona of market oscillations---cyclic thinning and expansion of Wi-Fi product adoption. Whenever sufficient backhaul bandwidth is present, i.e., whenever Wi-Fi represents the bottleneck, the market may be self-balancing, i.e., optimal utilization of available spectrum, or may on the other hand culminate in a system deadlock due to overutilization (tragedy of commons) or underutilization (insufficient Wi-Fi customer stock). On the contrary, whenever Wi-Fi capacity outperforms the backhaul high-speed access, no tragedy of commons is inferred, i.e., Wi-Fi is in capacity abundance.

\section*{Acknowledgments}

The Competence Center FTW Forschungszentrum Telekommunikation Wien GmbH is funded within the program COMET -- Competence Centers for Excellent Technologies by BMVIT, BMWA, and the City of Vienna. The COMET program is managed by the FFG. The authors would like to thank all members of the AWARE team for their input and for many fruitful discussions during the past years.


\bibliographystyle{IEEEtran}
\balance
\bibliography{lit}

\begin{thebibliography}{10}
\providecommand{\url}[1]{#1}
\csname url@samestyle\endcsname
\providecommand{\newblock}{\relax}
\providecommand{\bibinfo}[2]{#2}
\providecommand{\BIBentrySTDinterwordspacing}{\spaceskip=0pt\relax}
\providecommand{\BIBentryALTinterwordstretchfactor}{4}
\providecommand{\BIBentryALTinterwordspacing}{\spaceskip=\fontdimen2\font plus
\BIBentryALTinterwordstretchfactor\fontdimen3\font minus
  \fontdimen4\font\relax}
\providecommand{\BIBforeignlanguage}[2]{{%
\expandafter\ifx\csname l@#1\endcsname\relax
\typeout{** WARNING: IEEEtran.bst: No hyphenation pattern has been}%
\typeout{** loaded for the language `#1'. Using the pattern for}%
\typeout{** the default language instead.}%
\else
\language=\csname l@#1\endcsname
\fi
#2}}
\providecommand{\BIBdecl}{\relax}
\BIBdecl

\bibitem{FuxjagerFGR10}
P.~Fuxj{\"a}ger, H.~R. Fischer, I.~Gojmerac, and P.~Reichl, ``{Radio resource
  allocation in urban femto-WiFi convergence scenarios},'' in
  \emph{{Proceedings of the 6th EURO-NF Conference on Next Generation Internet
  (NGI)}}, 2010, pp. 1--8.

\bibitem{FuxjagerGFR11}
P.~Fuxj{\"a}ger, I.~Gojmerac, H.~R. Fischer, and P.~Reichl,
  ``{Measurement-Based Small-Cell Coverage Analysis for Urban Macro-Offload
  Scenarios},'' in \emph{{Proceedings of the 73rd IEEE Vehicular Technology
  Conference, VTC Spring}}, 2011, pp. 1--5.

\bibitem{han10}
B.~Han, P.~Hui, and A.~Srinivasan, ``{Mobile data offloading in metropolitan
  area networks},'' \emph{{ACM SIGMOBILE Mobile Computing and Communications
  Review}}, vol.~14, no.~4, 2010.

\bibitem{lee10}
K.~Lee, J.~Lee, Y.~Yi, I.~Rhee, and S.~Chong, ``{Mobile data offloading: how
  much can WiFi deliver?}'' in \emph{{Proceedings of the 6th International
  Conference on emerging Networking EXperiments and Technologies
  (CoNEXT'10)}}.\hskip 1em plus 0.5em minus 0.4em\relax {ACM}, 2010.

\bibitem{liu12}
S.~Liu and A.~Striegel, ``{Casting doubts on the viability of WiFi
  offloading},'' in \emph{{Proceedings of the 2012 ACM SIGCOMM workshop on
  Cellular networks: operations, challenges, and future design
  (CellNet'12)}}.\hskip 1em plus 0.5em minus 0.4em\relax {ACM}, 2012, pp.
  25--30.

\bibitem{siris13}
V.~A. Siris and D.~Kalyvas, ``{Enhancing Mobile Data Offloading with Mobility
  Prediction and Prefetching},'' in \emph{{Proceedings of the seventh ACM
  international workshop on Mobility in the evolving internet architecture
  (MobiArch'12)}}.\hskip 1em plus 0.5em minus 0.4em\relax {ACM}, 2013, pp.
  17--22.

\bibitem{casey10}
T.~Casey, T.~Smura, and A.~Sorri, ``{Vlaue Network COnfigurations in Wireless
  Local Area Access},'' in \emph{{Proceedings of the 9th Conference on
  Telecommunications Internet and Media Techno Economcis (CTTE)}}, 2010.

\bibitem{casey12}
T.~R. Casey and J.~T{\"o}yli, ``{Dynamics of two-sided platform success and
  failure: An analysis of public wireless local area access},''
  \emph{{Technovation}}, vol.~32, pp. 703--716, 2012.

\bibitem{gronsund13}
P.~Gr{\o}nsund, O.~Gr{\o}ndalen, and M.~L{\"a}hteenoja, ``{Business case
  evaluations for LTE network offloading with cognitive femtocells},''
  \emph{{Telecommunications Policy}}, vol.~37, no. 2--3, pp. 140--153,
  April--March 2013.

\bibitem{sigcomm13}
P.~Reichl, P.~Maill{\'e}, P.~Zwickl, and A.~Sackl, ``{A Fixed-Point Model for
  QoE-based Charging},'' in \emph{{ACM SIGCOMM 2013 Workshop on Future
  Human-Centric Multimedia Networking}}, {to appear 2013}.

\bibitem{shetty2008modeling}
N.~Shetty, G.~Schwartz, and J.~Walrand, ``{Modeling the Impact of QoS on
  Internet User Welfare and ISP Incentives},'' \emph{Working Paper, Berkeley
  University}, 2008.

\bibitem{Gibbens:2001ty}
R.~Gibbens, R.~Mason, and R.~Steinberg, ``{Internet Service Classes Under
  Competition},'' \emph{IEEE Journal on Selcted Areas in Communications},
  vol.~18, no.~12, pp. 2490--2498, 2001.

\end{thebibliography}

%



\end{document}